\documentclass[twocolumn,prX,nofootinbib]{revtex4-1}

\usepackage{amsmath,amssymb,amsfonts,amsthm,array,eucal,mathrsfs,upgreek,stackrel,slashed,graphicx,tensor,color,enumerate}
\usepackage{hyperref}

\definecolor{brown}{rgb}{0.65,0.16,0.16}

\newcommand{\beql}[1]{\begin{equation}\label{#1}}
\newcommand{\eeq}{\end{equation}}
\newcommand{\ie}{\textit{i.e.}~}

\newcommand{\dl}{\partial}
\newcommand{\Dl}{\nabla}

\newcommand{\udotchi}{\left(u \cdot \chi \right)}
\newcommand{\sdotchi}{\left(s \cdot \chi \right)}
\newcommand{\adotchi}{\left(a \cdot \chi \right)}
\newcommand{\adots}{\left(a \cdot s \right)}
\newcommand{\lieD}[2]{\mathsterling_{#1} #2}
\newcommand{\ed}{\mathrm{d}}

\newcommand{\dT}{\mathrm{d}T }

\newcommand{\smet}{\mathsf{h}}
\newcommand{\met}{\mathsf{g}}
\newcommand{\CC}{\textsc{cc}}

\newcommand{\lag}{\mathscr{L}}

\newcommand{\ruh}{r_\textsc{uh}}
\begin{document}
\author{David Mattingly}\email{david.mattingly@unh.edu}
\author{Matthew Roberson}\email{mkrbrson@gmail.com}
\affiliation{Department of Physics, University of New Hampshire, Durham, NH 03824, USA}

\title{Algebra of universal horizon preserving diffeomorphisms}
\author{Sayandeb Basu}\email{sayandeb@honors.usf.edu}
\affiliation{University of South Florida, Tampa, FL 33620}



\begin{abstract}
In relativistic gravity, requiring a spacetime hypersurface be a Killing horizon breaks the general covariance of general relativity.  The residual algebra of horizon preserving diffeomorphisms can be extended to a Virasoro algebra near the horizon, the central charge of which yields the Bekenstein-Hawking entropy via the Cardy formula.  This near horizon symmetry approach provides an argument for why black hole entropy computations in various quantum gravity models all agree.  An exception may be Ho\v{r}ava-Lifshitz gravity, where causal horizons are not Killing horizons but rather universal horizons.  As a first step towards determining if the entropy of universal horizons can be calculated by a near horizon symmetry approach we compute the classical algebra of universal horizon preserving diffeomorphisms.  We find that the algebra is similar to the algebra in the Killing horizon case, but with a modification that agrees with other approaches to universal horizon thermodynamics.    
\end{abstract}

\maketitle


\section{Introduction}
Despite vast differences in their underlying microscopic descriptions, various approaches to quantum gravity yield the same black hole entropy - independent of charge, spin, horizon topology, and number of spacetime dimensions -  given by the Bekenstein-Hawking formula \cite{Carlip:2014pma}. The ubiquity of this result led to the argument by Strominger \cite{Strominger:1997eq} and Carlip \cite{Carlip:1998wz} that a classically inherited near horizon symmetry forces any quantum gravity states to admit a conformal field theory description. This conformal controlling symmetry at the horizon dictates the ultra-violet density of states and hence the black hole entropy, which can be calculated via the Cardy formula \cite{Cardy:1986ie}. This symmetry construction holds for any generic black hole without regard for the quantum gravitational details, thus providing an apparent explanation for the universality of black hole entropy \cite{Carlip:2011vr} intimately related to conformal field theory.  

Black hole entropy led to holography, and conformal techniques remain a key ingredient in AdS/CFT.  Many physical systems, such as non-relativistic condensed matter systems, are not however described by CFTs. Finding successful non-relativistic extensions of holographic methods is therefore a necessary step towards applying holographic techniques to a vastly larger class of systems.  Steps along this path have been taken by examining gravitational duals to Lifshitz field theories.  Any gravitational dual to a Lifshitz field theory must admit solutions that exhibit Lifshitz symmetry somewhere in the spacetime. Since global Lifshitz geometry is not a solution of the vacuum Einstein equations, most relativistic duals of Lifshitz field theories contain extra matter fields~\cite{Kachru:2008yh, Goldstein:2009cv, Taylor:2008tg} with non-trivial bulk profiles. 

An exception to this is Ho\v{r}ava-Lifshitz gravity, a renormalizable theory of quantum gravity with a Lifshitz symmetry in the ultraviolet~\cite{Horava:2009uw}.  Lifshitz spacetimes are vacuum solutions of the field equations and, in fact, are much more natural than AdS which only occurs in a particular corner of the parameter space. Ho\v{r}ava-Lifshitz gravity further holds promise for Lifshitz holography\cite{Griffin:2012qx, Cheyne:2017bis}.  Since black hole entropy has been such a key component of holography, it is therefore important to understand how to reconcile the near horizon symmetry approach, which forces any quantum gravity theory to admit a CFT description of microscopic near horizon states, with the seemingly incompatible inherent high energy Lifshitz symmetry of Ho\v{r}ava-Lifshitz gravity.

A way out of this conflict is provided by the nature of black hole solutions in Ho\v{r}ava-Lifshitz gravity, which are very different than horizons in relativistic gravity theories and, as we shall see, possess a different near horizon symmetry.  Ho\v{r}ava-Lifshitz gravity assumes that there exists a preferred foliation of spacetime, which allows one to break Lorentz invariance at high energies and impose the Lifshitz symmetry.   Despite the violation of Lorentz invariance, which permits superluminal excitations, black hole solutions exist\cite{Blas:2011ni,Eling:2006ec}.  However, the static causal horizons in these solutions, called ''universal horizons''~\cite{Blas:2011ni}, are distinct from Killing horizons, as they must be due to the presence of superluminal excitations.  This is very unlike the situation in general relativity where the two horizons coincide.  Known solutions have been found to possess universal horizons which obey the first law of black hole mechanics and have been argued to emit radiation \cite{Berglund:2012bu, Berglund:2012fk, Basu:2016vyz, Ding:2015fyx, Wang:2017brl}.

Implementation of the near horizon symmetry approach relies on a particular construction~\cite{Carlip:1999cy}: require a horizon be present, argue that the relevant quantum gravity states are those annihilated by this condition, identify the corresponding algebra of horizon preserving diffeomorphisms, extend it to a Virasoro algebra, and count the states via the Cardy formula~\cite{Cardy:1986ie}. The key question addressed in this article is the first step - finding the classical near-horizon algebra of universal horizon preserving diffeomorphisms given the assumption that a static universal horizon is present.  This is a step towards a possible quantum symmetry and relationship between universal horizon entropy and Lifshitz extensions of the Cardy formula.  The paper is organized as follows. In section~\ref{HL}, we introduce Ho\v{r}ava gravity, its action, and some important notes regarding its causal structure and black hole solutions. In section~\ref{NH} we review the near horizon symmetry construction of black hole entropy for static horizons in general relativity and construct the corresponding symmetry for universal horizons. Finally, we summarize in the conclusions~\ref{Conclusions}.


\section{Ho\v{r}ava-Lifshitz Background}\label{HL}
\subsection{Ho\v{r}ava Gravity}
Ho\v{r}ava gravity is a presumed renormalizable quantum field theory of gravity based off ideas from the theory of quantum critical phenomena.  In Ho\v{r}ava gravity general relativity is an effective low energy limit of a theory with Lifshitz symmetry in the ultraviolet, which in turn requires that spacetime comes equipped with a preferred foliation. The existence of this foliation allows one to introduce higher-order spatial derivatives into the action while maintaining second-order in time derivatives; this makes the theory power-counting renormalizable without introducing ghosts and violating unitarity \cite{Stelle:1976gc, Visser:2009fg,Visser:2009ys, Sotiriou:2010wn}. The cost of introducing a preferred foliation is a violation of Lorentz symmetry in the UV, but there are schemes that allow for the recovery of Lorentz invariance in the infrared so that the theory is compatible with experiment~\cite{Pospelov:2010mp}.

If we employ an ADM decomposition for the metric in the preferred foliation,
\begin{equation}
ds^2=-N^2 dt^2 + \smet_{ij}(dx^i + N^i dt)(dx^j + N^j dt),
\end{equation}
then the action for Ho\v{r}ava-Lifshitz gravity takes the form
\cite{Sotiriou:2011dr, Blas:2009qj}
\begin{equation*}\label{hl:action:general}
S_{HL} = \frac{M_{Pl}^2}{2} \int \dT d^3 x N \sqrt{\smet} \left(\lag_2 + \frac{\lag_4}{M^2_{*}} + \frac{\lag_6 }{M^4_{*}}+ \cdots\right)~,
\end{equation*}
where $M_{Pl}$ is the Planck mass $M_{Pl} = (8 \pi G)^{-1/2}$, $\smet$ is the determinant of the induced metric $\smet_{ij}$ on the spacelike hypersurfaces, and
\begin{equation*}\label{hl:action:L2}
\lag_2 = K_{ij} K^{ij} - \lambda K^2 + \xi \left(^{(3)}R - 2 \Lambda \right) + \eta a_i a^i~,
\end{equation*}
with $\lambda$, $\xi$, $\eta$ dimensionless coupling constants. $K$ is the trace of the extrinsic curvature $K_{ij}$, $^{(3)}R$ is the Ricci scalar of $\smet_{ij}$, $N$ is the lapse function, and $a_i = \dl_i\ln{N}$.

$\lag_4$ and $\lag_6$ and above are collections of 4th, 6th, etc. order operators and $M_{*}$ is the scale that suppresses these operators. $M_{*}$ does not coincide \emph{a priori} with the Planck mass, and in fact must be less than $M_{PL}$ if the theory is not to become strongly coupled near the Planck scale~\cite{Papazoglou:2009fj}. It is these higher dimension operators which encode the Lifshitz symmetry of the theory in the ultraviolet.  The Lifshitz symmetry, which is after all a non-relativistic symmetry, allows high frequency modes to propagate arbitrarily fast with respect to a leaf of the preferred foliation.  Note that this is perfectly causal; even though high energy modes propagate faster than the speed of light closed causal curves and other pathologies are still forbidden.

We usually ignore any possible higher dimension irrelevant operators when finding horizons in general relativity.  For example, for stationary spacetimes we find a Killing horizon via solving the equations of motion generated by the infrared, two derivative Ricci scalar term.  Implicitly, however, when we promote a Killing horizon to a causal horizon we assume that local Lorentz invariance holds exactly, so that even high frequency modes are trapped behind the Killing horizon.  A similar approach holds here.  A geometric definition of the universal horizon and its location in a static spacetime can be solved for using the infrared sector of the theory, but promoting it to a causal boundary requires a statement about the ultraviolet sector. We expand a bit on the infrared sector of Ho\v{r}ava-Lifshitz below, and then introduce black holes and causal horizons in the next section.

There is a convenient, covariant formulation of Ho\v{r}ava gravity as a scalar-tensor theory, where a dynamical scalar field $T$, called the~\emph{khronon}, labels the leaves of the foliation.  Since a preferred foliation is a requirement, $T$ always admits a non-zero timelike gradient everywhere. This allows one to construct a unit-timelike hypersurface orthogonal one-form $u_a$, called the~\emph{{\ae}ther}, such that
\beql{ae:HSO:norm}
u_a = -N\Dl_a T, \qquad \met^{a b}u_a u_b = -1~,
\eeq
where the lapse $N$ is now solved for via the unit norm constraint by 
\beql{eq:lapse}
N^{-2} = -\met^{a b}(\Dl_a T)(\Dl_b T)~.
\eeq
Besides the usual diffeomorphisms, Ho\v{r}ava gravity must also be invariant under arbitrary reparametrizations of the khronon: $T \mapsto \tilde{T} = \tilde{T}(T)$, as the existence of the foliation does not depend on the labeling. Under such reparametrizations  $N$ is required to transform as $N \mapsto \tilde{N} = (\ed\tilde{T}/\dT)^{-1}N$, such that the {\ae}ther remains manifestly invariant under the reparametrizations of the khronon. 

This above re-formulation allows one to express the (two-derivative truncated/IR limit) action of Ho\v{r}ava gravity not in the ADM language, but in a manifestly covariant and reparametrization invariant manner as follows~\cite{Jacobson:2010mx}\footnote{The complete action of Ho\v{r}ava gravity can also be covariantized via such a `St\"uckelberg' procedure~\cite{Sotiriou:2011dr}. In this work, however, we only work with the IR limit of the theory.}
\beql{action}
S = \frac{1}{16\pi G_{\ae}}\int\ed^3x\sqrt{-\met}(-2\Lambda_{\CC} + R + \lag).
\eeq
Here $\Lambda_{\CC}$ is the cosmological constant which will be taken to be negative in this work, $R$ is the curvature scalar, and $\lag$ is the khronon's Lagrangian  given by
\beql{lag:ae}
\lag = -\tensor{Z}{^{a b}_{c d}}(\Dl_a u^c)(\Dl_b u^d)~.
\eeq
The tensor $\tensor{Z}{^{a c}_{c d}}$ is given by
\beql{def:Zabcd}
\tensor{Z}{^{a b}_{c d}} = c_1\met^{a b}\met_{c d} + c_2\tensor{\delta}{^a_c}\tensor{\delta}{^b_d} + c_3\tensor{\delta}{^a_d}\tensor{\delta}{^b_c} - c_4u^au^b\met_{c d}~,
\eeq
where $c_1$, $c_2$, $c_3$, $c_4$ are coupling constants.  The bulk covariant equations of motion for the metric and khronon are generated by extremizing the action~\eqref{action} under variations of the respective fields, with the assumption that the {\ae}ther is derived from the khronon via~\eqref{ae:HSO:norm}.  With this notation and background in hand, we now turn to black holes and horizons in Ho\v{r}ava-Lifshitz gravity.

\subsection{Black Holes in Ho\v{r}ava Gravity}
Black holes in general relativity, with its foundational assumption of local Lorentz symmetry, are characterized by the presence of an event horizon: a causal boundary which traps the fastest possible excitations allowed by the theory. Since Ho\v{r}ava gravity exhibits a non-relativistic Lifshitz symmetry in the UV one might conclude that such a causal boundary - and hence black holes - cannot exist. However, in recent years, black hole solutions of Ho\v{r}ava gravity have been discovered and explored - chiefly in highly symmetric and/or static spacetimes (see, for example, \cite{Barausse:2011pu, Sotiriou:2014gna, Bhattacharyya:2014kta, Ding:2015fyx,Wang:2017brl}). 

\begin{figure}[htb]
	\centering
	\includegraphics[scale=0.7]{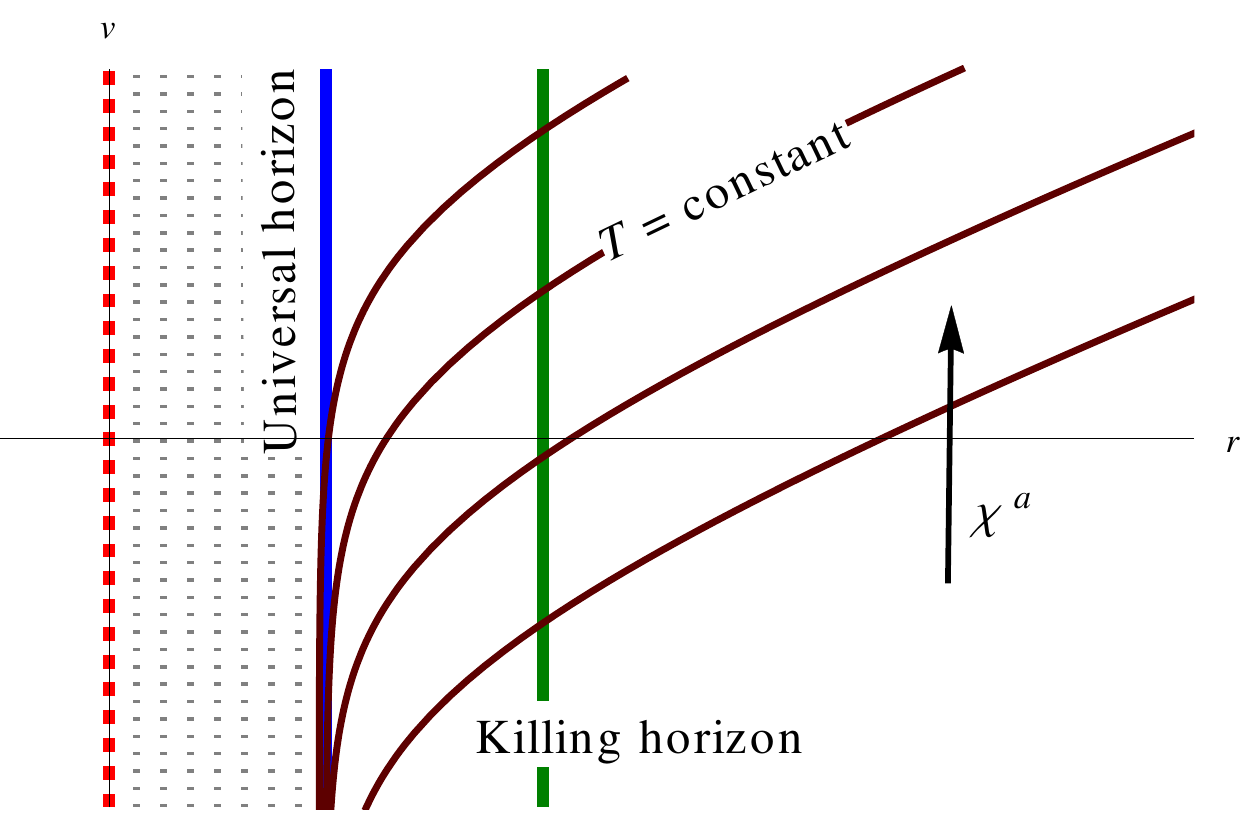}
	\caption{Bending of the preferred ($T = $ constant) hypersurfaces (thick brown lines) in ingoing Eddington-Finkelstein type coordinates in a static and spherically symmetric black hole solution of Ho\v{r}ava gravity. The Killing vector $\chi^a$ points upward throughout everywhere. The vertical green line is a constant $r$ hypersurface and denotes the usual Killing horizon defined by $\chi^a\chi^b \met_{a b} = 0$. The universal horizon of a Ho\v{r}ava gravity black hole, denoted by the vertical blue line, is also a constant $r$ hypersurface (located at $r = \ruh$) defined by the condition $u_a\chi^a = 0$, where $u_a$ is the unit timelike normal vector to the constant $T$ hypersurfaces. The dotted region inside the universal horizon (\ie for $r < \ruh$) denotes a black hole region even for arbitrary fast excitations; the constant $T$ hypersurfaces for this region are not shown to keep the diagram clean.}
	\label{fig:bendingT}
\end{figure}

The causal structure of (generally covariant) Ho\v{r}ava gravity - inherited from the foliation structure of the spacetime manifold required for UV completion and retention of unitarity - plays an important role in the definition of a black hole spacetime. The scalar khronon field $T$ always has an everywhere-timelike gradient whose level sets foliate the spacetime with spacelike hypersurfaces. Because of the theory's Lifshitz symmetry in the UV, ultra-high frequency modes can travel arbitrarily fast. However, these excitations always move 'forward in time' with respect to the preferred foliation and are not required to be contained within any propagation cone. Hence these constant-$T$ hypersurfaces, the leaves of the foliation, impose a preferred notion of simultaneity and endow the theory with a truly non-relativistic causal structure.\footnote{This differs from spacetimes with only a preferred frame - such as those which arise in the more general Lorentz-violating Einstein-{\AE}ther gravitational theory - which have a general-relativity-like 'quasi-relativistic' causal structure. } In black hole solutions the leaves of the foliation bend inwards as one moves inwards from spatial infinity, as in figure \ref{fig:bendingT}. However, one of these leaves - the constant$/(T \rightarrow \infty)$ hypersurface - does not reach spatial infinity. This hypersurface is the \emph{universal horizon}. For a general discussion of the causal structure of spacetimes in theories with a preferred foliation see~\cite{Bhattacharyya:2015gwa}.  For our purposes we need only the local definition of non-extremal universal horizons in stationary spacetimes. These are characterized by the conditions
\begin{equation}
\left(u \cdot \chi\right) = 0~, \hspace{0.5cm} \left(a \cdot \chi\right) = \text{constant} \neq 0,
\end{equation} 
where $\chi^a$ is the Killing vector which generates constant-r hypersurfaces. Since the khronon field (and thus the {\ae}ther field) defines an absolute time, any excitation trapped inside the universal horizon must necessarily move 'backward in time' with respect to the preferred foliation in order to escape to infinity, violating the theory's non-relativistic notion of causality. Thus, respecting causality, the universal horizon acts as a true causal horizon trapping all excitations of arbitrary speed.\newline

Universal horizons have been found to obey a first law \cite{Berglund:2012bu, Mohd:2013zca}. Furthermore, tunneling calculations using eternal universal horizon geometries indicate that they radiate thermally \cite{Berglund:2012fk, Cropp:2013sea, Ding:2015fyx}. In both cases the role of surface gravity/temperature is played by $a \cdot s$, where $s^a$ (which shall be introduced in detail later) is a unit spacelike vector that becomes parallel to $\chi$ at the universal horizon. Calculations in collapsing geometries give a different picture \cite{Michel:2015rsa} when considering low frequency emission.  Additionally, universal horizons may be unstable in certain regions of parameter space~\cite{Blas:2011ni}.  In short, the status of thermodynamics of universal horizons is very much up in the air.  In the below we examine the near horizon diffeomorphisms under the philosophy that \textit{if} a thermodynamics holds, is there a near-horizon symmetry argument that can in principle be applied to calculate the entropy?  And if so, does $a \cdot s$ play the role of the surface gravity in this case as well?
  
\section{Near-Horizon Symmetry Constructions}\label{NH}

\subsection{Construction for Killing horizons in general relativity}
The universality of black hole thermodynamics has motivated a search for a general principle governing the underlying statistical mechanics that is independent of the details of any particular quantum theory of gravity. A major step in finding such a principle, based on symmetries, was taken by Strominger \cite{Strominger:1997eq}. Previous work by Brown and Henneaux \cite{Brown:1986nw} showed that the diffeomorphism invariance of gravity in $\text{AdS}_3$ can be manifested as a two-dimensional CFT "living" at the boundary of the spacetime. With this in mind Strominger observed, in the context of (2+1)-dimensional BTZ black holes, that symmetry arguments can be used to count states in quantum gravity via the Cardy formula, which determines the asymptotic density of states entirely in terms of the Virasoro algebra and its central charge - independent of other details of the theory. He showed that using the central charge in Cardy's formula reproduces the standard Bekenstein-Hawking entropy, and thus proposed this symmetry construction as a possible resolution to the universality problem. 

Subsequent work by Carlip \cite{Carlip:1998wz, Carlip:1999cy, Carlip:2002be} and Solodukhin \cite{Solodukhin:1998tc} modified and extended this work to arbitrary-dimensional spacetimes and symmetries near the horizon rather than at infinity. The CFT in this program exists at the black hole horizon, rather than at the boundary of the spacetime.  These studies showed that the algebra of horizon preserving diffeomorphisms also was a Virasoro algebra with a central charge that generated the Bekenstein-Hawking entropy appropriately.  We now detail this latter construction.

Carlip proposed a general recipe for building a so-called ``black hole/CFT correspondence'' based on the observation that, while details vary, most arguments for such dualities incorporate similar steps. This general algorithm, given in  {\cite{Carlip:2011vr} and \cite{Carlip:2012ff}, can be described as follows:
\begin{enumerate}
	\item Identify an appropriate boundary and impose corresponding boundary conditions specifying the horizon properties of the black hole.
	\item Determine how these boundary conditions affect the symmetries of general relativity, i.e. the allowed diffeomorphisms and associated surface deformations.
	\item Seek a preferred classical subalgebra of diffeomorphisms (e.g., $\textit{Diff}\left(S^1\right)$ or $\textit{Diff}\left(S^1\right) \times \textit{Diff}\left(S^1\right)$). If such an algebra exists, promote it to a Virasoro algebra.
	\item Read off the central charges and count the states using the Cardy formula.
\end{enumerate}

Stationary universal horizons are most similar in terms of their geometric construction to stationary Killing horizons.  We will therefore summarize the application of the first three steps above for Killing horizons~\cite{Carlip:1999cy}, before showing how to apply the algorithm to (stationary) universal horizons. Since in this paper we are only interested in identifying the corresponding algebra of diffeomorphisms and not in the calculation of black hole entropy via the Cardy formula, we postpone the fourth step to a future publication. 

The first three steps of Carlip's algorithm (i.e., those required for finding a preferred subalgebra of diffeomorphisms) for the case of a generic Killing horizon can be summarized as follows. (See \cite{Carlip:1999cy} for Carlip's original work.)
\subsubsection{Impose Killing Horizon Boundary Conditions}
\indent Consider an $N$-dimensional region $\mathscr{M}$ of a spacetime $\left(\mathscr{N}, g_{ab}\right)$ with a Killing vector $\chi^a$ and a Killing horizon characterized by $\chi^2 = \chi^a \chi^b g_{a b} = 0$ at some surface $S$ in $\mathscr{M}$. We now require that $S$ remains a Killing horizon under diffeomorphisms that act on the dynamical objects (in this case $g_{ab}$) and that the near horizon asymptotic structure remains unchanged.  The diffeomorphisms must then satisfy the  following constraints:
\begin{equation}\label{eq:KH:Constraint1}
\frac{\delta\left(\chi^2\right)}{\chi^2} = \frac{\chi^a \chi^b \delta g_{ab}}{\chi^2} \rightarrow 0 \hspace{0.5cm}  \text{as } \chi^2 \rightarrow 0
\end{equation} 
\begin{equation}\label{eq:KH:Constraint2}
\text{and} \hspace{0.5cm} \chi^a t^b \delta g_{a b} \rightarrow 0 \hspace{0.5cm}  \text{as } \chi^2 \rightarrow 0 ~,
\end{equation} 
where $t^a$ is any unit spacelike vector tangent to $S$ (i.e. they are spacelike generators of the Killing horizon). These constraints force the relative variation in the norm of $\chi^a$ to vanish and the tangent vectors on the horizon to remain orthogonal, preserving the surface as a Killing horizon. Note that, in practice for general relativity, these constraints must be put on a stretched horizon, where $\chi^2 = \epsilon$ and the limit $\epsilon \rightarrow 0$ is taken at the end of the calculation.

\subsubsection{Find the boundary preserving diffeomophisms}
Next, the diffeomorphisms which preserve the horizon condition above is sought. Near the horizon, define a vector field $\rho^a$ which is everywhere orthogonal to $\chi^a$ by 
\begin{equation} \label{KH:rhoDef}
\nabla_a \chi^2  = -2 \kappa \rho_a~,
\end{equation}
where $\kappa$ is the surface gravity at the horizon. 
For a diffeomorphism generated by a vector field $\xi^a$, the above boundary conditions \eqref{eq:KH:Constraint1} and \eqref{eq:KH:Constraint2}, with $\delta g_{a b} = \mathsterling_{\xi} g_{ab} = \nabla_a \xi_b + \nabla_b \xi_a$, put restrictions on the corresponding allowed diffeomorphisms, implying that $\xi^a$ must satisfy 
\begin{equation}
\frac{\chi^a \chi^b}{\chi^2} \nabla_a \xi_b = \chi^a \nabla_a \left(\frac{\chi_b \xi^b}{\chi^2}\right) - \kappa \frac{\rho_b \xi^b}{\chi^2} = 0~.
\end{equation} 
If we restrict as usual to diffeomorphisms in the $r$-$t$ plane~\cite{Dreyer:2013noa}, since there is reason to believe that black hole entropy in general relativity is effectively described by a 2-d conformal field theory ~\cite{Carlip:2006kq, 
Birmingham:2001qa}
then $\xi^a$ is of the form 
\begin{equation}\label{eq:KH:Diffeo}
\xi^a = T \chi^a + R \rho^a~,
\end{equation}
where $T$ and $R$ are functions of spacetime and do not necessarily satisfy any Killing symmetry. In principle, this is a two parameter family. However, the impositions of the boundary conditions above enforces a relationship between $R$ and $T$ 
\begin{equation}\label{eq:KH:CoefficientRelation}
R = \frac{1}{\kappa} \frac{\chi^2}{\rho^2} \nabla_{\chi} T~,
\end{equation}
where $\nabla_v \equiv v^a \nabla_a$ denotes the directional derivative in the direction of some vector field $v^a$.  This results in a one parameter family.

\subsubsection{Find a Preferred Subalgebra}
\indent The diffeomorphisms defined by \eqref{eq:KH:Diffeo} and \eqref{eq:KH:CoefficientRelation}  do not form a closed algebra - one must add additional conditions, which we now do. To obtain closure, it is required that if $\xi_1^a,\xi_2^a$ are two vector fields that satisfy \eqref{eq:KH:Diffeo} and \eqref{eq:KH:CoefficientRelation} then if we define $\xi_3^a$ by $ \xi_3^a=\left\{\xi_1, \xi_2 \right\}^a \equiv \mathsterling_{\xi_1} \xi^a_2$ 
it must be of the form $\xi_3^a= T_3 \chi^a + \left(\frac{1}{\kappa} \frac{\chi^2}{\rho^2} \nabla_{\chi} T_3 \right) \rho^a$.\newline
Computing the commutator yields
\begin{equation} \label{eq:KH:Commutator}
	\left\{ \xi_1, \xi_2 \right\}^a = \mathsterling_{\xi_1} \xi^a_2
	= T_3 \chi^a + R_3\rho^a~,
\end{equation}
with
\begin{align}\label{eq:KH:Tfound}
	T_3 &= \left[T_1, T_2\right]_{\nabla_{\chi}} + \frac{1}{\kappa} \frac{\chi^2}{\rho^2} \left(\left(\nabla_{\chi} T_1\right)\left(\nabla_{\rho} T_2\right) - \left(\nabla_{\chi} T_2\right)\left(\nabla_{\rho} T_1\right)\right)
\end{align} 
and 
\begin{align}\label{eq:KH:Rfound}
	R_3 &= \frac{1}{\kappa} \frac{\chi^2}{\rho^2} \nabla_{\chi} \left(\left[T_1, T_2\right]_{\nabla_{\chi}}\right) + \frac{1}{\kappa^2} \frac{\chi^4}{\rho^4} \left[\nabla_{\chi} T_1, \nabla_{\chi} T_2\right]_{\nabla_{\rho}}~,
\end{align} 
where we define $\left[f_1, f_2\right]_D \equiv f_1 D f_2 - f_2 D f_1$, for scalar functions $f_1$ and $f_2$ and an appropriate differential operator $D$.

Computing 
\begin{equation}\label{eq:KH:generalizedClosure}
	\Delta R_3 \equiv R_3 - \frac{1}{\kappa} \frac{\chi^2}{\rho^2} \nabla_{\chi} T_3~,
\end{equation}
and setting \eqref{eq:KH:generalizedClosure} equal to zero results in the following condition required for closure:
\begin{equation} \label{eq:KH:closureCondition}
	\rho^a \nabla_a T = 0 \hspace{0.5cm} \text{for } n\ =\ 1,\ 2~,
\end{equation}
for all vector fields $\xi$ at the Killing horizon. Applying the condition \eqref{eq:KH:closureCondition} to the commutator \eqref{eq:KH:Commutator} therefore closes the algebra. The vector field $\xi_3^a = \left\{\xi_1, \xi_2 \right\}^a$ can then be written explicitly as 
\begin{equation}\label{eq:KH:closedSubalgebra}
	\left\{\xi_1, \xi_2 \right\}^a = \left(\left[T_1, T_2\right]_{\nabla_{\chi}}\right) \chi^a + \frac{1}{\kappa} \frac{\chi^2}{\rho^2} \nabla_{\chi} \left(\left[T_1, T_2\right]_{\nabla_{\chi}}\right) \rho^a~.
\end{equation}

\indent Now that a preferred subalgebra \eqref{eq:KH:closedSubalgebra} has been found, it remains to be identified.  This is done in the Killing horizon case by appealing to Euclidean approaches to black hole thermodynamics.  The Hawking temperature of a black hole can be derived by requiring that the corresponding Euclidean black hole is periodic in imaginary time $\tau$ with periodicity $2\pi/\kappa$.  This periodicity is leveraged to restrict the sub-algebra above to periodic functions in the following way. Let $\mathscr{H}$ denote the $(N-2)-$dimensional intersection of a Cauchy surface $\mathscr{C}$ with the Killing horizon. We Fourier decompose $T$ on the Killing horizon as
\begin{equation}
T_n (v, \theta^i) = \frac{1}{\kappa} e^{i n \kappa v} f_n\left(\theta^i\right)
\end{equation}
so that $T = \sum_{n} A_n T_n$, where $v$ denotes a parameter along the orbits of the Killing vector $\chi^a$ (normalized so that $\nabla_{\chi} v = 1$), $\theta^i$ are angular coordinates on $\mathscr{H}$, and $f_m\left(\theta^i\right)f_n\left(\theta^i\right) = f_{m+n}\left(\theta^i\right)$ by spherical symmetry. Then, by \eqref{eq:KH:closedSubalgebra}, the following relation is satisfied:
\begin{equation}\label{KH:diffS1}
\left\{\xi_m, \xi_n\right\}^a = -i (m-n) \xi_{m+n}^a~.
\end{equation}
This is the well known Witt algebra $\textit{Diff}\left(S^1\right)$ \cite{Schottenloher:2008zz}. Hence, the subalgebra of diffeomorphism generators given by \eqref{eq:KH:closedSubalgebra} is isomorphic to $\textit{Diff}\left(S^1\right)$.  This algebra can then be promoted to a Virasoro algebra.  For this last part of step three and step four, the central extension and counting of states, we again refer the reader to~\cite{Carlip:1999cy} as we shall find that there is an obstruction to implementing these steps in the universal horizon case.

\subsection{Construction for universal horizons in Ho\v{r}ava-Lifshitz gravity}\label{NUH}
We now turn to identifying the near horizon symmetries for a static universal horizon, in exact analogy to the Killing horizon construction above.  
\subsubsection{Universal Horizon Boundary Conditions}
\indent Consider an $N$-dimensional region $\mathscr{M}$ of a foliated spacetime $\left(\mathscr{N}, \Sigma, g_{ab}\right)$, with foliation structure $\Sigma$. $\Sigma$ is the set of all leaves of the preferred foliation. We label the leaves of the preferred foliation by the scalar khronon function $T$ and denote one such leaf by $\Sigma_T$. Each leaf is a constant khronon hypersurface and acts as a surface of simultaneity. Furthermore let $\mathscr{M}$ admit a universal horizon characterized by $u \cdot \chi  = 0$ on some surface $S$, where $u^a$ is the unit timelike normal vector to the constant $T$ hypersurfaces and $\chi^a$ is a Killing vector.  Note that $\chi^a$ is a Killing vector for \textit{both} $g_{ab}$ and $u_a$, i.e.  $\mathsterling_{\chi} g_{ab} = \mathsterling_{\chi} u_a = 0$. We now make a similar horizon-preserving ansatz to that of the Killing horizon case. Consider diffeomorphisms which preserve the location of the universal horizon and its asymptotic structure. Fix $\chi^a$ and take the diffeomorphisms to act on $u_a,g_{ab}$, since they are the relevant dynamical fields. The corresponding horizon preserving constraints are therefore
\begin{equation}\label{eq:UH:boundaryCondition1}
\frac{\delta\left(u \cdot \chi\right)}{\left(u \cdot \chi\right)} = \frac{\chi^a \delta u_a}{\left(u \cdot \chi\right)} \rightarrow 0 \hspace{0.5cm}  \text{as } \left(u \cdot \chi\right) \rightarrow 0
\end{equation} 
\begin{equation}\label{eq:UH:boundaryCondition2}
\text{and} \hspace{0.5cm} \chi^a t^b \delta g_{a b} \rightarrow 0 \hspace{0.5cm}  \text{as } \left(u \cdot \chi\right) \rightarrow 0 ~,
\end{equation} 
where $t^a$ is any unit spacelike vector tangent to $\partial \mathscr{M}$. These constraints force the relative variation of $\left(u \cdot \chi\right)$ to vanish and the tangent vectors on the horizon to remain orthogonal, preserving the surface as a universal horizon.

\subsubsection{Find the boundary preserving algebra of diffeomorphisms}
We first need to construct a set of basis vectors that span the $r$-$t$ plane, similar to $\chi^a$ and $\rho^a$ in the Killing horizon case. The simplest choice of basis is the $\{u^a,\ s^a\}$ basis, where $s^a$ is the outward-pointing vector tangent to the foliation hypersurfaces defined by $s^2 = s_a s^a = 1$ and $\left(u \cdot s\right) = u_a s^a = 0$. For a diffeomorphism generated by a vector field $\xi^a$, the above boundary conditions \ref{eq:UH:boundaryCondition1} and \ref{eq:UH:boundaryCondition2}, with $\delta u_a = \mathsterling_{\xi} u_a = \xi^b \nabla_b u_a + u_b \nabla_a \xi^b$, imply that $\xi^a$ must satisfy 
\begin{equation}
\xi^a \nabla_a \left(u \cdot \chi \right) + \chi^a \nabla_a \left(u \cdot \xi\right) = 0
\end{equation}
in the near horizon region.  If we assume a spherically symmetric spacetime (for simplicity),\footnote{In a spherically symmetric spacetime, the Killing vector $\chi^a$ can be expanded in the $\{u^a, s^a\}$ basis as $\chi^a = -\left(u \cdot \chi\right) u^a + \left(s \cdot \chi\right) s^a$.} restrict our attention to the $r$-$t$ plane, and expand $\xi^a$ in the $\{u^a, s^a\}$ basis so that
\begin{equation}\label{eq:UH:Diffeo}
\xi^a = U u^a + S s^a~,
\end{equation}
where $U$ and $S$ are functions of spacetime and do not necessarily satisfy any Killing symmetry, then the boundary condition \ref{eq:KH:Constraint1} provides a relation between $U$ and $S$ given by
\begin{equation}\label{eq:UH:CoefficientRelation}
S = \frac{\nabla_{\chi} U}{\left(a \cdot s\right)\left(u \cdot \chi\right)}  - \frac{\left(s \cdot \chi\right)}{\left(u \cdot \chi\right)} U~,
\end{equation}
which is a one parameter family. \newline

\subsubsection{Finding a Preferred Subalgebra}
\indent Next, the conditions under which the diffeomorphisms characterized by \eqref{eq:UH:Diffeo} and \eqref{eq:UH:CoefficientRelation} form a closed subalgebra are to be determined. To obtain closure, it is required that 
\begin{equation} \label{eq:closurereq}
\xi_3^a=\left\{\xi_1, \xi_2 \right\}^a  = U_3 u^a + \left(\frac{\nabla_{\chi} U_3}{\adots \udotchi} - \frac{\sdotchi}{\udotchi} U_3\right) s^a
\end{equation} 
for some diffeomorphism-generating vector fields $\xi_1^a$, $\xi_2^a$, $\xi_3^a$ satisfying \eqref{eq:UH:Diffeo} and \eqref{eq:UH:CoefficientRelation}.
The computation of the commutator is rather tedious and unenlightening and hence has been relegated to Appendix~\ref{sec:AppendixA}.  The result is
\begin{equation} \label{eq:UH:Commutator}
\left\{ \xi_1, \xi_2 \right\}^a
= U_3 u^a + S_3 s^a~,
\end{equation}
with
\begin{align}\label{eq:UH:Ufound}
U_3 &= \frac{-1}{\adots \udotchi} \left(\left(\nabla_s U_1\right)\left(\nabla_{\chi} U_2\right)-\left(\nabla_s U_2\right)\left(\nabla_{\chi} U_1\right)\right)
\end{align} 
and 
\begin{align}\label{eq:UH:Sfound}
S_3 &= \nabla_{\chi} \left(\frac{-1}{\adots \udotchi^2} \left[ U_1, U_2 \right]_{\nabla_{\chi}}\right) \nonumber \\ 
&\hspace{0.5cm} + \frac{1}{\adots^2 \udotchi^2} \left[\nabla_{\chi} U_1, \nabla_{\chi} U_2 \right]_{\nabla_{s}} \nonumber \\ 
&\hspace{0.5cm} + \frac{\sdotchi}{\adots \udotchi^2} \left(\left(\nabla_s U_1\right)\left(\nabla_{\chi} U_2\right)-\left(\nabla_s U_2\right)\left(\nabla_{\chi} U_1\right)\right)~.
\end{align} 

If this algebra is to be closed, then $S_3$ must match the coefficient of $s^a$ in ~\eqref{eq:closurereq}. Computing 
\begin{equation}\label{eq:UH:generalizedClosure}
\Delta S_3 \equiv S_3 - \left(\frac{\nabla_{\chi} U_3}{\adots \udotchi} - \frac{\sdotchi}{\udotchi} U_3\right)~,
\end{equation}
and setting $\Delta S_3$ equal to zero results in the following condition required for closure:
\begin{equation}\label{eq:UH:closureCondition}
\nabla_s U_n - \adots U_n = 0
\end{equation}
which is the analog of~\eqref{eq:KH:closureCondition}.  The explicit details of this calculation can again be found in the Appendix. The vector field $\xi_3^a = \left\{\xi_1, \xi_2 \right\}^a$ can then be written explicitly as
\begin{equation}\label{eq:UH:closedSubalgebra}
\left\{\xi_1, \xi_2 \right\}^a = \left(\left[U_1, U_2\right]_{\nabla_u}\right)u^a + \left(\frac{-1}{\adots} \nabla_u \left(\left[U_1, U_2\right]_{\nabla_u}\right)\right) s^a~
\end{equation}
which is the primary result of this work.

\section{From $\textit{Diff}(R)$ to $\textit{Diff}(S)$?}
The similarity of \eqref{eq:UH:closedSubalgebra} and \eqref{eq:KH:closedSubalgebra} is clear and certainly suggestive, in that the quantity $a \cdot s$ in the UH case plays the role of $\kappa$ in the KH case.  $a \cdot s$ has been found to control the temperature of UH in the tunneling approach~\cite{Berglund:2012fk} as well as the first law~\cite{Berglund:2012bu,Basu:2016vyz}.  Continuing the path towards $\textit{Diff}(S^1)$ and \eqref{KH:diffS1} is, however, problematic in the UH case with Lifshitz fields.  The algebras found in \eqref{eq:KH:closedSubalgebra} and \eqref{eq:UH:closedSubalgebra} are representations of $\textit{Diff}(R^1)$ i.e. diffeomorphisms of the real line.  This is to be expected, since once transverse diffeomorphisms are ignored both the KH and UH are simply embeddings of $R^1$ into the black hole spacetime.

The motivation for examining $\textit{Diff}(S^1)$ , i.e. \eqref{KH:diffS1}, in the Killing horizon case relies on arguments from Euclideanization of the black hole spacetime in general relativity (c.f. the discussion in~\cite{Ross:2005sc}).  However no Euclidean solution for universal horizons yet exists. And, important field theory questions such as the extension of the Hadamard condition to Lifshitz fields and whether regularity on the universal horizon preferentially selects a state equivalent to the Hartle-Hawking state are almost completely unexplored.  We leave these questions for future work.

What we can say is that there is no \textit{mathematical} obstruction to finding a $\textit{Diff}(S^1)$ algebra.  For example, the universal horizon is a leaf of the foliation for which the Killing vector is a generator.  Let us denote by $\mu$ a parameter generated by the orbits of $u^a$ normalized so that $\nabla_{u} \mu = 1$. We now Fourier decompose the coefficient function $U$ such that
\begin{equation}
U_n (\mu, \theta^i) = \frac{1}{\adotchi} e^{i n \adotchi \mu} f_n\left(\theta^i\right)
\end{equation}
so that $U = \sum_{n} A_n U_n$, where $\theta^i$ are again transverse coordinates on $\mathscr{H}$, and $f_m\left(\theta^i\right)f_n\left(\theta^i\right) = f_{m+n}\left(\theta^i\right)$. Then, by \eqref{eq:UH:closedSubalgebra}, the following relation is satisfied:
\begin{equation}\label{UH:diffS1}
\left\{\xi_m, \xi_n\right\}^a = -i (m-n) \xi_{m+n}^a~.
\end{equation}
This is the algebra of $\textit{Diff}\left(S^1\right)$. We stress again, however, that whether such subalgebras have any physical meaning is unknown.



\section{Conclusion}\label{Conclusions}
Since Ho\v{r}ava-Lifshitz gravity can be formulated as a covariant theory in the infrared, the algebra of diffeomorphisms is the same as in general relativity.  Requiring a black hole horizon be present therefore can, in principle, be used to calculate the entropy via a near-horizon symmetry argument.  Whether a near-horizon symmetry approach can work in Ho\v{r}ava-Lifshitz gravity is an important question, as the near-horizon approach has been used to argue for the universality of black hole entropy independent of quantum gravity.  In Ho\v{r}ava-Lifshitz gravity, however, the causal horizons in static spacetimes are not Killing horizons, but universal horizons.  These universal horizons have a different geometrical definition, i.e. $u \cdot \chi=0$ instead of $\chi \cdot \chi=0$, and hence may provide an exception to the universal argument. 

A key aspect of the near-horizon symmetry approach is whether there exists a closed sub-algebra of horizon preserving diffeomorphisms.  In this work, we showed that a closed sub-algebra does indeed exist and moreover its algebraic structure is very similar to the Killing horizon case from general relativity.  As well, the role of the surface gravity again appears to be played by the quantity $a \cdot s$, which is consistent with past tunneling and mechanics results on the possible thermodynamics of universal horizons.  Additionally one can further restrict this algebra to the classical Witt algebra as in the general relativistic/Killing horizon case.  However, unlike the Killing horizon case there is no physical motivation to do so.  Hence while indications are promising that the near-horizon symmetry approach can be applied to universal horizons as well, more work needs to be done, especially on applying Euclidean techniques to universal horizons, to provide the correct physical motivation and guidance necessary for the next steps in the program.

\begin{acknowledgments}
DM and MR thank the University of New Hampshire for research support.

\end{acknowledgments}

\appendix
\clearpage
\section{Calculational Details}\label{sec:AppendixA}
\subsection{Commutators in $u^a, s^a$ basis}
	The commutator ${\xi_1,\xi_2}^a$ in the $u^a, s^a$ basis with $\xi_n=U_n u^a + S_n s^a$, is
\begin{align} \label{eq:UH:closure1}
\left\{\xi_1, \xi_2 \right\}^a &= \mathsterling_{\xi_1} \xi^a_2 = \xi^b_1 \nabla_b \xi^a_2 - \xi^b_2 \nabla_b \xi^a_1 \nonumber \\
&= \left(\left[U_1, U_2\right]_{\nabla_{u}} + \left(S_1 \nabla_{s} U_2 - S_2 \nabla_{s} U_1\right)\right) u^a \nonumber \\ 
&\hspace{0.5cm} + \left(\left(U_1 \nabla_{u} S_2 - U_2 \nabla_{u} S_1\right) + \left[S_1, S_2\right]_{\nabla_{s}}\right) s^a \nonumber \\
&\hspace{0.5cm} + \left(U_1 S_2 - U_2 S_1\right) \mathsterling_{u} s^a~.
\end{align} 

\noindent By~\eqref{eq:UH:CoefficientRelation} the following relations between the various directional derivatives of $U_1,U_2,S_1,S_2$ hold
\begin{align*}
&\left(S_1 \nabla_{s} U_2 - S_2 \nabla_{s} U_1\right) = \hspace{5cm} \\
 &  \frac{1}{\left(a \cdot s\right) \left(u \cdot \chi\right)} \left(\left(\nabla_{\chi} U_1\right)\left(\nabla_{s} U_2\right) - \left(\nabla_{\chi} U_2\right)\left(\nabla_{s} U_1\right)\right)\nonumber \\
&  -\frac{\left(s \cdot \chi\right)}{\left(u \cdot \chi\right)} \left[U_1, U_2\right]_{\nabla_{s}}~,
\end{align*}
\begin{align*}
&\left(U_1 \nabla_{u} S_2 - U_2 \nabla_{u} S_1\right) = \hspace{5cm}\\
&\nabla_u \left(\frac{1}{\left(a \cdot s\right)\left(u \cdot \chi\right)}\right) \left[U_1, U_2\right]_{\nabla_{\chi}}  \\
&+ \frac{1}{\left(a \cdot s\right)\left(u \cdot \chi\right)} \left(U_1 \nabla_{u} \left(\nabla_{\chi} U_2\right) - U_2 \nabla_u \left(\nabla_{\chi} U_1\right)\right) \nonumber \\
& - \frac{\left(s \cdot \chi\right)}{\left(u \cdot \chi\right)} \left[U_1, U_2\right]_{\nabla_u}~,
\end{align*}
\begin{align*}
&\left[S_1, S_2\right]_{\nabla_{\chi}} = \\
&\frac{1}{\left(a \cdot s\right)^2 \left(u \cdot \chi\right)^2} \left[\nabla_{\chi} U_1, \nabla_{\chi} U_2\right]_{\nabla_{s}}  \\
& - \frac{\left(s \cdot \chi \right)}{\left(a \cdot s\right) \left(u \cdot \chi\right)^2} \left(U_1 \nabla_s \left(\nabla_{\chi} U_2\right) - U_2 \nabla_s \left(\nabla_{\chi} U_1\right)\right) \nonumber \\
& + \frac{1}{\left(u \cdot \chi\right)^2} \left(\frac{\nabla_s \left(s \cdot \chi\right)}{\left(a \cdot s\right)} - \left(s \cdot \chi\right) \nabla_s \left(\frac{1}{\left(a \cdot s\right)}\right)\right)\left[U_1, U_2\right]_{\nabla_{\chi}} \nonumber \\
& - \frac{\left(s \cdot \chi\right)}{\left(a \cdot s\right)\left(u \cdot \chi\right)^2} \left(\left(\nabla_{\chi} U_1\right)\left(\nabla_s U_2\right) - \left(\nabla_{\chi} U_2\right)\left(\nabla_s U_1\right)\right) \nonumber \\
& + \frac{\left(s \cdot \chi\right)^2}{\left(u \cdot \chi\right)^2}\left[U_1, U_2\right]_{\nabla_s}~,
\end{align*}
and
\begin{align*}
\left(U_1 S_2 - U_2 S_1\right) = \frac{1}{\left(a \cdot s\right)\left(u \cdot \chi\right)} \left[U_1, U_2\right]_{\nabla_{\chi}}.
\end{align*}
Applying these relations our commutator \eqref{eq:UH:closure1} becomes
\begin{widetext}
\begin{align} \label{eq:UH:closure2}
\left\{\xi_1, \xi_2 \right\}^a &= \left(\left[U_1, U_2\right]_{\nabla_{u}} -\frac{\left(s \cdot \chi\right)}{\left(u \cdot \chi\right)} \left[U_1, U_2\right]_{\nabla_{s}} + \frac{1}{\left(a \cdot s\right) \left(u \cdot \chi\right)} \left(\left(\nabla_{\chi} U_1\right)\left(\nabla_{s} U_2\right) - \left(\nabla_{\chi} U_2\right)\left(\nabla_{s} U_1\right)\right)\right) u^a \nonumber \\
&\hspace{0.5cm} + \left(\nabla_u \left(\frac{1}{\left(a \cdot s\right)\left(u \cdot \chi\right)}\right) \left[U_1, U_2\right]_{\nabla_{\chi}} + \frac{1}{\left(a \cdot s\right)\left(u \cdot \chi\right)} \left(U_1 \nabla_{u} \left(\nabla_{\chi} U_2\right) - U_2 \nabla_u \left(\nabla_{\chi} U_1\right)\right) \right) s^a \nonumber \\
&\hspace{0.5cm} + \left(-\frac{\left(s \cdot \chi\right)}{\left(u \cdot \chi\right)} \left[U_1, U_2\right]_{\nabla_u} + \frac{1}{\left(a \cdot s\right)^2 \left(u \cdot \chi\right)^2} \left[\nabla_{\chi} U_1, \nabla_{\chi} U_2\right]_{\nabla_{s}}\right) s^a \nonumber \\
&\hspace{0.5cm} + \left(- \frac{\left(s \cdot \chi \right)}{\left(a \cdot s\right) \left(u \cdot \chi\right)^2} \left(U_1 \nabla_s \left(\nabla_{\chi} U_2\right) - U_2 \nabla_s \left(\nabla_{\chi} U_1\right)\right) \right) s^a \nonumber \\
&\hspace{0.5cm} + \left(\frac{1}{\left(u \cdot \chi\right)^2} \left(\frac{\nabla_s \left(s \cdot \chi\right)}{\left(a \cdot s\right)} - \left(s \cdot \chi\right) \nabla_s \left(\frac{1}{\left(a \cdot s\right)}\right)\right)\left[U_1, U_2\right]_{\nabla_{\chi}} \right) s^a \nonumber \\
&\hspace{0.5cm} + \left(-\frac{\left(s \cdot \chi\right)}{\left(a \cdot s\right)\left(u \cdot \chi\right)^2} \left(\left(\nabla_{\chi} U_1\right)\left(\nabla_s U_2\right) - \left(\nabla_{\chi} U_2\right)\left(\nabla_s U_1\right)\right) + \frac{\left(s \cdot \chi\right)^2}{\left(u \cdot \chi\right)^2}\left[U_1, U_2\right]_{\nabla_s} \right) s^a \nonumber \\
&\hspace{0.5cm} + \left(\frac{1}{\left(a \cdot s\right)\left(u \cdot \chi\right)} \left[U_1, U_2\right]_{\nabla_{\chi}}\right) \mathsterling_{u} s^a~.
\end{align}
This can be simplified to
\begin{align} \label{eq:UH:closure3}
\left\{\xi_1, \xi_2 \right\}^a &= \left(\frac{-1}{\left(u \cdot \chi \right)}\left[U_1, U_2\right]_{\nabla_{\chi}} + \frac{1}{\left(a \cdot s\right) \left(u \cdot \chi\right)} \left(\left(\nabla_{\chi} U_1\right)\left(\nabla_{s} U_2\right) - \left(\nabla_{\chi} U_2\right)\left(\nabla_{s} U_1\right)\right)\right) u^a \nonumber \\
&\hspace{0.5cm} + \left(\left(\frac{\nabla_s \left(s \cdot \chi\right)}{\left(a \cdot s\right) \left(u \cdot \chi\right)^2} + \frac{\nabla_{\chi} \left(a \cdot s \right)}{\left(a \cdot s\right)^2 \left(u \cdot \chi\right)^2}\right)\left[U_1, U_2\right]_{\nabla_{\chi}} + \frac{1}{\left(a \cdot s\right)^2 \left(u \cdot \chi\right)^2} \left[\nabla_{\chi} U_1, \nabla_{\chi} U_2\right]_{\nabla_s} \right) s^a \nonumber \\
&\hspace{0.5cm} + \left(\frac{-1}{\left(a \cdot s\right) \left(u \cdot \chi\right)^2} \nabla_{\chi} \left(\left[U_1, U_2\right]_{\nabla_{\chi}}\right) + \frac{\left(s \cdot \chi\right)}{\left(a \cdot s\right) \left(u \cdot \chi\right)^2} \left(\left(\nabla_s U_1\right)\left(\nabla_{\chi} U_2\right) - \left(\nabla_s U_2\right)\left(\nabla_{\chi} U_1\right)\right)\right) s^a \nonumber \\
&\hspace{0.5cm} + \left(\frac{1}{\left(a \cdot s\right)\left(u \cdot \chi\right)} \left[U_1, U_2\right]_{\nabla_{\chi}}\right) \mathsterling_{u} s^a~.
\end{align}
\end{widetext}
Expanding $\lieD{u}{s^a}$ yields $\lieD{u}{s^a} = \adots u^a - K_{ss} s^a$, where $K_{ss}$ is the extrinsic curvature of the foliation projected along $s^a s^b,$ i.e. $K_{ss}=s^a s^b \nabla_a u_b$ . This allows us to expand the above expression in terms of the $u^a,s^a$ basis as
\begin{widetext}
\begin{align} \label{eq:UH:closure4}
\left\{\xi_1, \xi_2 \right\}^a &= \left(\frac{-1}{\adots \udotchi} \left(\left(\nabla_s U_1\right)\left(\nabla_{\chi} U_2\right) - \left(\nabla_{s} U_2\right)\left(\nabla_{\chi} U_1\right)\right)\right) u^a \nonumber \\
&\hspace{0.5cm} + \left( \nabla_{\chi}\left(\frac{-1}{\adots \udotchi^2} \left[U_m, U_n\right]_{\nabla_{\chi}}\right) + \frac{1}{\adots^2 \udotchi^2} \left[\nabla_{\chi} U_m, \nabla_{\chi} U_n\right]_{\nabla_s}\right) s^a \nonumber \\
&\hspace{0.5cm} + \left(\frac{\sdotchi}{\adots \udotchi^2} \left(\left(\nabla_s U_m\right)\left(\nabla_{\chi} U_n\right)-\left(\nabla_s U_n\right)\left(\nabla_{\chi} U_m\right)\right)\right) s^a.
\end{align}
\end{widetext}
The coefficients in~\eqref{eq:UH:Ufound} and~\eqref{eq:UH:Sfound} can then be directly read off from the expression above.

\subsection{Closure relations}
\noindent We first define
\[\left\{\xi_1, \xi_2 \right\}^a = \left(U_3\right)_{\text{found}} u^a + \left(S_3\right)_{\text{found}} s^a\] 
with $\left(U_3\right)_{\text{found}}$ and $ \left(S_3\right)_{\text{found}}$ defined as the $u^a, s^a$ coefficients in~\eqref{eq:UH:closure4}.
In order for the algebra to close $S_3$ must satisfy ~\eqref{eq:UH:CoefficientRelation}, i.e.
\begin{equation*}
\left\{\xi_1, \xi_2 \right\}^a = U_3 u^a + S_3 s^a = U_3 u^a + \left(S_3\right)_{\text{needed}} s^a~,
\end{equation*} 
with \[\left(U_3\right)_{\text{found}} = U_3\] and 
\begin{widetext}
\begin{align*}
\left(S_3\right)_{\text{needed}} &= \frac{-1}{\adots \udotchi} \nabla_{\chi} U_3 - \frac{\sdotchi}{\udotchi} U_3\\
&= - \frac{1}{\adots^2 \udotchi^2} \left(\left(\nabla_s U_m\right)\nabla_{\chi}\left(\nabla_{\chi} U_n\right)-\left(\nabla_s U_n\right)\nabla_{\chi}\left(\nabla_{\chi} U_m\right)\right) \nonumber \\ 
&\hspace{0.5cm} - \frac{1}{\adots^2 \udotchi^2} \left(\left(\nabla_{\chi} U_m\right)\nabla_{\chi}\left(\nabla_{s} U_n\right)-\left(\nabla_{\chi} U_n\right)\nabla_{\chi}\left(\nabla_{s} U_m\right)\right) \nonumber \\ 
&\hspace{0.5cm} + \frac{\sdotchi}{\adots \udotchi^2} \left(\left(\nabla_s U_m\right)\left(\nabla_{\chi} U_n\right)-\left(\nabla_s U_n\right)\left(\nabla_{\chi} U_m\right)\right).
\end{align*} 
Closure is equivalent to the statement

\begin{align*}
0 &= \Delta S_3 = \left(S_3\right)_{\text{found}} - \left(S_3\right)_{\text{needed}} \\ &= \frac{1}{\adots^2 \udotchi^2} \left(\left(\nabla_s U_m - \adots U_m\right)\nabla_{\chi}\left(\nabla_{\chi} U_n\right)-\left(\nabla_s U_n - \adots U_n\right)\nabla_{\chi}\left(\nabla_{\chi} U_m\right)\right)~,
\end{align*}
\end{widetext}
which is ensured by the condition
\begin{equation}
\nabla_s U - \adots U = 0~.
\end{equation}
By applying this closure condition to \eqref{eq:UH:Commutator}, noting that $\nabla_{\chi} = -(u\cdot \chi) \nabla_{u} + (s \cdot \chi) \nabla_{s}$, the vector field $\xi_3^a = \left\{\xi_1, \xi_2 \right\}^a$ can then be written explicitly as
\begin{equation}
\left\{\xi_1, \xi_2 \right\}^a = \left(\left[U_m, U_n\right]_{\nabla_u}\right) + \left(\frac{-1}{\adots} \nabla_u \left(\left[U_m, U_n\right]_{\nabla_u}\right)\right) s^a~.
\end{equation}


\begin{thebibliography}{99}
\bibitem{Carlip:2014pma} 
S.~Carlip,
Int.\ J.\ Mod.\ Phys.\ D {\bf 23}, 1430023 (2014)
doi:10.1142/S0218271814300237
[arXiv:1410.1486 [gr-qc]].


\bibitem{Strominger:1997eq} 
A.~Strominger,
JHEP {\bf 9802}, 009 (1998)
doi:10.1088/1126-6708/1998/02/009
[hep-th/9712251].


\bibitem{Carlip:1998wz} 
S.~Carlip,
Phys.\ Rev.\ Lett.\  {\bf 82}, 2828 (1999)
doi:10.1103/PhysRevLett.82.2828
[hep-th/9812013].


\bibitem{Carlip:2011vr} 
S.~Carlip,
Entropy {\bf 13}, 1355 (2011)
doi:10.3390/e13071355
[arXiv:1107.2678 [gr-qc]].


\bibitem{Kachru:2008yh} 
S.~Kachru, X.~Liu and M.~Mulligan,
Phys.\ Rev.\ D {\bf 78}, 106005 (2008)
doi:10.1103/PhysRevD.78.106005
[arXiv:0808.1725 [hep-th]].


\bibitem{Goldstein:2009cv} 
K.~Goldstein, S.~Kachru, S.~Prakash and S.~P.~Trivedi,
JHEP {\bf 1008}, 078 (2010)
doi:10.1007/JHEP08(2010)078
[arXiv:0911.3586 [hep-th]].


\bibitem{Taylor:2008tg} 
M.~Taylor,
arXiv:0812.0530 [hep-th].


\bibitem{Horava:2009uw} 
P.~Ho\v{r}ava,
Phys.\ Rev.\ D {\bf 79}, 084008 (2009)
doi:10.1103/PhysRevD.79.084008
[arXiv:0901.3775 [hep-th]].

\bibitem{Griffin:2012qx} 
T.~Griffin, P.~Hořava and C.~M.~Melby-Thompson,
Phys.\ Rev.\ Lett.\  {\bf 110}, no. 8, 081602 (2013)
doi:10.1103/PhysRevLett.110.081602
[arXiv:1211.4872 [hep-th]].

\bibitem{Cheyne:2017bis} 
J.~Cheyne and D.~Mattingly,
Phys.\ Rev.\ D {\bf 97}, no. 6, 066024 (2018)
doi:10.1103/PhysRevD.97.066024
[arXiv:1707.05913 [gr-qc]].


\bibitem{Eling:2006ec} 
C.~Eling and T.~Jacobson,
Class.\ Quant.\ Grav.\  {\bf 23}, 5643 (2006)
Erratum: [Class.\ Quant.\ Grav.\  {\bf 27}, 049802 (2010)]
doi:10.1088/0264-9381/23/18/009, 10.1088/0264-9381/27/4/049802
[gr-qc/0604088].

\bibitem{Blas:2011ni} 
D.~Blas and S.~Sibiryakov,
Phys.\ Rev.\ D {\bf 84}, 124043 (2011)
doi:10.1103/PhysRevD.84.124043
[arXiv:1110.2195 [hep-th]].

\bibitem{Berglund:2012bu} 
P.~Berglund, J.~Bhattacharyya and D.~Mattingly,
Phys.\ Rev.\ D {\bf 85}, 124019 (2012)
doi:10.1103/PhysRevD.85.124019
[arXiv:1202.4497 [hep-th]].


\bibitem{Berglund:2012fk} 
P.~Berglund, J.~Bhattacharyya and D.~Mattingly,
Phys.\ Rev.\ Lett.\  {\bf 110}, no. 7, 071301 (2013)
doi:10.1103/PhysRevLett.110.071301
[arXiv:1210.4940 [hep-th]].


\bibitem{Basu:2016vyz} 
S.~Basu, J.~Bhattacharyya, D.~Mattingly and M.~Roberson,
Phys.\ Rev.\ D {\bf 93}, no. 6, 064072 (2016)
doi:10.1103/PhysRevD.93.064072
[arXiv:1601.03274 [hep-th]].

\bibitem{Ding:2015fyx} 
C.~Ding, A.~Wang, X.~Wang and T.~Zhu,
Nucl.\ Phys.\ B {\bf 913}, 694 (2016)
doi:10.1016/j.nuclphysb.2016.10.007
[arXiv:1512.01900 [gr-qc]].

\bibitem{Wang:2017brl} 
A.~Wang,
Int.\ J.\ Mod.\ Phys.\ D {\bf 26}, no. 07, 1730014 (2017)
doi:10.1142/S0218271817300142
[arXiv:1701.06087 [gr-qc]].

\bibitem{Carlip:1999cy} 
S.~Carlip,
Class.\ Quant.\ Grav.\  {\bf 16}, 3327 (1999)
doi:10.1088/0264-9381/16/10/322
[gr-qc/9906126].

\bibitem{Cardy:1986ie} 
J.~L.~Cardy,
Nucl.\ Phys.\ B {\bf 270}, 186 (1986).
doi:10.1016/0550-3213(86)90552-3


\bibitem{Stelle:1976gc} 
K.~S.~Stelle,
Phys.\ Rev.\ D {\bf 16}, 953 (1977).
doi:10.1103/PhysRevD.16.953

\bibitem{Visser:2009fg} 
M.~Visser,
Phys.\ Rev.\ D {\bf 80}, 025011 (2009)
doi:10.1103/PhysRevD.80.025011
[arXiv:0902.0590 [hep-th]].

\bibitem{Visser:2009ys} 
M.~Visser,
arXiv:0912.4757 [hep-th].

\bibitem{Sotiriou:2010wn} 
T.~P.~Sotiriou,
J.\ Phys.\ Conf.\ Ser.\  {\bf 283}, 012034 (2011)
doi:10.1088/1742-6596/283/1/012034
[arXiv:1010.3218 [hep-th]].

\bibitem{Pospelov:2010mp} 
M.~Pospelov and Y.~Shang,
Phys.\ Rev.\ D {\bf 85}, 105001 (2012)
doi:10.1103/PhysRevD.85.105001
[arXiv:1010.5249 [hep-th]].

\bibitem{Sotiriou:2011dr} 
T.~P.~Sotiriou, M.~Visser and S.~Weinfurtner,
Phys.\ Rev.\ D {\bf 83}, 124021 (2011)
doi:10.1103/PhysRevD.83.124021
[arXiv:1103.3013 [hep-th]].

\bibitem{Blas:2009qj} 
D.~Blas, O.~Pujolas and S.~Sibiryakov,
Phys.\ Rev.\ Lett.\  {\bf 104}, 181302 (2010)
doi:10.1103/PhysRevLett.104.181302
[arXiv:0909.3525 [hep-th]].

\bibitem{Papazoglou:2009fj} 
A.~Papazoglou and T.~P.~Sotiriou,
Phys.\ Lett.\ B {\bf 685}, 197 (2010)
doi:10.1016/j.physletb.2010.01.054
[arXiv:0911.1299 [hep-th]].

\bibitem{Jacobson:2010mx} 
T.~Jacobson,
Phys.\ Rev.\ D {\bf 81}, 101502 (2010)
Erratum: [Phys.\ Rev.\ D {\bf 82}, 129901 (2010)]
doi:10.1103/PhysRevD.82.129901, 10.1103/PhysRevD.81.101502
[arXiv:1001.4823 [hep-th]].

\bibitem{Barausse:2011pu} 
E.~Barausse, T.~Jacobson and T.~P.~Sotiriou,
Phys.\ Rev.\ D {\bf 83}, 124043 (2011)
doi:10.1103/PhysRevD.83.124043
[arXiv:1104.2889 [gr-qc]].


\bibitem{Sotiriou:2014gna} 
T.~P.~Sotiriou, I.~Vega and D.~Vernieri,
Phys.\ Rev.\ D {\bf 90}, no. 4, 044046 (2014)
doi:10.1103/PhysRevD.90.044046
[arXiv:1405.3715 [gr-qc]].


\bibitem{Bhattacharyya:2014kta} 
J.~Bhattacharyya and D.~Mattingly,
Int.\ J.\ Mod.\ Phys.\ D {\bf 23}, no. 13, 1443005 (2014)
doi:10.1142/S0218271814430056
[arXiv:1408.6479 [hep-th]].

\bibitem{Bhattacharyya:2015gwa} 
J.~Bhattacharyya, M.~Colombo and T.~P.~Sotiriou,
Class.\ Quant.\ Grav.\  {\bf 33}, no. 23, 235003 (2016)
doi:10.1088/0264-9381/33/23/235003
[arXiv:1509.01558 [gr-qc]].

\bibitem{Mohd:2013zca} 
A.~Mohd,
arXiv:1309.0907 [gr-qc].

\bibitem{Cropp:2013sea} 
B.~Cropp, S.~Liberati, A.~Mohd and M.~Visser,
Phys.\ Rev.\ D {\bf 89}, no. 6, 064061 (2014)
doi:10.1103/PhysRevD.89.064061
[arXiv:1312.0405 [gr-qc]].

\bibitem{Michel:2015rsa} 
F.~Michel and R.~Parentani,
Phys.\ Rev.\ D {\bf 91}, no. 12, 124049 (2015)
doi:10.1103/PhysRevD.91.124049
[arXiv:1505.00332 [gr-qc]].


\bibitem{Brown:1986nw} 
J.~D.~Brown and M.~Henneaux,
Commun.\ Math.\ Phys.\  {\bf 104}, 207 (1986).
doi:10.1007/BF01211590

\bibitem{Carlip:2002be} 
S.~Carlip,
Phys.\ Rev.\ Lett.\  {\bf 88}, 241301 (2002)
doi:10.1103/PhysRevLett.88.241301
[gr-qc/0203001].


\bibitem{Solodukhin:1998tc} 
S.~N.~Solodukhin,
Phys.\ Lett.\ B {\bf 454}, 213 (1999)
doi:10.1016/S0370-2693(99)00398-6
[hep-th/9812056].

\bibitem{Carlip:2012ff} 
S.~Carlip,
AIP Conf.\ Proc.\  {\bf 1483}, 54 (2012)
doi:10.1063/1.4756962
[arXiv:1207.1488 [gr-qc]].

\bibitem{Dreyer:2013noa} 
O.~Dreyer, A.~Ghosh and A.~Ghosh,
Phys.\ Rev.\ D {\bf 89}, no. 2, 024035 (2014)
doi:10.1103/PhysRevD.89.024035
[arXiv:1306.5063 [gr-qc]].

\bibitem{Carlip:2006kq} 
S.~Carlip,
J.\ Phys.\ Conf.\ Ser.\  {\bf 33}, 73 (2006).
doi:10.1088/1742-6596/33/1/007

\bibitem{Birmingham:2001qa} 
D.~Birmingham, K.~S.~Gupta and S.~Sen,
Phys.\ Lett.\ B {\bf 505}, 191 (2001)
doi:10.1016/S0370-2693(01)00354-9
[hep-th/0102051].


\bibitem{Schottenloher:2008zz} 
M.~Schottenloher,
Lect.\ Notes Phys.\  {\bf 759}, 1 (2008).
doi:10.1007/978-3-540-68628-6

\bibitem{Ross:2005sc} 
S.~F.~Ross,
hep-th/0502195.





\end{thebibliography}
\end{document}